\documentstyle[aaspp4]{article}

\begin{document}

\title{Statistical properties of SGR 1806-20 bursts}

\author{Ersin {G\"o\u{g}\"u\c{s}}\altaffilmark{1,3}, 
Peter M. Woods\altaffilmark{1,3}, Chryssa Kouveliotou\altaffilmark{2,3},
Jan van Paradijs\altaffilmark{1,4}, Michael S. Briggs\altaffilmark{1,3},
Robert C. Duncan\altaffilmark{5}, Christopher Thompson\altaffilmark{6}}

\altaffiltext{1}{Department of Physics, University of Alabama in Huntsville,
       Huntsville, AL 35899} 
\altaffiltext{2}{Universities Space Research Association}
\altaffiltext{3}{NASA Marshall Space Flight Center, SD-50, Huntsville, AL
35812}
\altaffiltext{4}{Astronomical Institute ``Anton Pannekoek'', University of
      Amsterdam, 403 Kruislaan, 1098 SJ Amsterdam, NL} 
\altaffiltext{5}{Department of Astronomy, University of Texas, RLM 15.308, 
      Austin, TX 78712-1083}
\altaffiltext{6}{Department of Physics and Astronomy, University of North
      Carolina, Philips Hall, Chapel Hill, NC, 27599-3255}

\authoremail{Ersin.Gogus@msfc.nasa.gov}

\begin{abstract}

We present statistics of SGR 1806-20 bursts, combining
290 events detected with 
RXTE/PCA, 111 events detected with BATSE and 134 events detected with ICE.  
We find that the fluence distribution of bursts 
observed with each instrument 
are well described  
by power laws with indices 1.43, 1.76 and 1.67, respectively.
The distribution of time intervals between successive bursts from 
SGR 1806-20 is described by a lognormal function with a peak at 103 s. 
There is no 
correlation between the burst intensity 
and either the waiting times till the next burst or the time elapsed
since the previous burst.  
In all these statistical properties, SGR 1806-20 bursts resemble 
a self-organized critical system, similar to earthquakes 
and solar flares. 
Our results thus support the hypothesis that the energy source for SGR bursts 
is crustquakes due to the evolving, strong magnetic field of the neutron star, 
rather than any accretion or nuclear power.

\end{abstract}


\section{Introduction}

Soft gamma repeaters (SGR) are a rare class of objects 
characterized by their repetitive emission of low energy gamma-ray bursts. 
SGR bursts last $\sim$ 0.1 s and their spectra are usually well
described by an optically thin thermal bremsstrahlung (OTTB) model with 
kT $\sim$ 20--40 keV. Three of the four known SGRs are
associated with slowly rotating (P$_{\rm spin}$ $\sim$ ~5--8 s; 
Mazets et al. 1979, Kouveliotou et al. 1998, Hurley et al. 1999), 
ultra-strongly magnetized ($B \gtrsim 10^{14}$ Gauss; 
Kouveliotou et al. 1998, Kouveliotou et al. 1999a) neutron stars
positioned within or near young supernova remnants. 
For a review of the burst and persistent emission properties of SGRs, 
see Kouveliotou (1999b) and Hurley (2000).

Cheng et al. (1996) \markcite{cheng96} reported similarities
between particular statistical properties of a sample of 111 SGR 1806-20 bursts
(observed with the International Cometary Explorer, ICE, between 1979 and 1984) 
and
earthquakes. They noted that the distribution of the event energies of both 
phenomena follow a
power law, dN $\propto$ E$^{-\gamma}$~dE, with index, $\gamma$ $\sim$
1.6. Furthermore, they found that the cumulative waiting times between 
successive SGR 
bursts and earthquakes are similar.
Laros et al. (1987) noted that the distribution of waiting times between 
successive SGR 1806-20 bursts follow a lognormal function, which was also seen 
between micro-glitches of the Vela pulsar
(Hurley et al. 1994). Using the same data set, Palmer (1999) showed that, 
similar to
earthquakes, some SGR 1806-20 bursts may originate from relaxation systems. 
{G\"o\u{g}\"u\c{s}} et al. (1999) studied a set of 1024 bursts from SGR
1900+14;
187 bursts were detected with the Burst and Transient Source Experiment (BATSE)
aboard the Compton Gamma Ray Observatory (CGRO)
and 837 bursts were detected with the Proportional Counter Array (PCA)
on the Rossi X-ray Timing Explorer (RXTE)
during an active period of the source in 1998. 
We found that their fluence distribution is consistent 
with a power law of index $\gamma$ = 1.66 over 4 orders of magnitude.
The distribution of waiting times between 
successive bursts also follows a lognormal function, which peaks at 
$\sim$ 49 s. 
We discussed the idea that SGRs, like earthquakes and solar flares, 
are manifestations
of self-organized critical systems (Bak, Tang \& Wiesenfeld 1988).
All of these results are consistent with
the idea that SGR bursts are caused by starquakes,
which are the result of a fracture of the crust of a magnetically-powered 
neutron star, 
or ``magnetar" (Duncan \& Thompson 1992\markcite{dt92}; Thompson and Duncan
1995\markcite{td95}, 1996\markcite{td96}).

SGR 1806-20 exhibited sporadic bursting activity from the launch of BATSE 
(in April 1991) until November 1993 (Kouveliotou et al. ~1994\markcite{kou94}).
In October 1996, the source entered a burst active phase. 
The reactivation initiated a series 
of pointed observations with the RXTE/PCA over a period of two weeks.
These observations led to the discovery of 7.47 s pulsations from SGR 1806-20
and confirmed its nature as a magnetar (Kouveliotou et al. 1998).  
In these two weeks RXTE/PCA recorded a total of 290 bursts\setcounter{footnote}{6}
\footnote{Examples of  RXTE/PCA observations of SGR 1806-20 can be
seen at {\tt http://gammaray.msfc.nasa.gov/batse/sgr/sgr1806/}}.
In the BATSE data, SGR 1806-20 burst activity was persistent but variable from
October 1996 up to October 1999 with a total of 116 recorded bursts.
In this {\it Letter}, we present a comprehensive study of the statistical 
properties of SGR 1806-20 by combining several data bases. Sections 2, 3 and 4
describe the CGRO/BATSE, RXTE/PCA and ICE observations, respectively. 
Our results are
presented in Section 5 and discussed in Section 6.

\section{BATSE Observations} 

In our analysis we
have used DISCriminator Large Area detector (DISCLA) data with coarse 
energy resolution (4
channels covering energies from 25 keV to $\sim$2 MeV), Spectroscopy 
Time-Tagged Event (STTE) data and
Spectroscopy High Energy Resolution Burst (SHERB) data with fine energy
binning (256 channels covering energies from 15 keV to $\sim$10 MeV) 
from the Spectroscopy Detectors. 
A detailed description of BATSE instrumentation
and data types can be found in Fishman et al.~(1989)\markcite{fish89}.

BATSE triggered on 74 bursts between September 1993 and
June 1999. For 32 of the brightest events, STTE or SHERB data
with detailed spectral information were obtained. 
The background subtracted spectra were fit to optically-thin thermal
bremsstrahlung
(OTTB) and power law models. The OTTB model, F(E)$\propto$ 
E$^{-1}$$\exp$($-$E/kT), provided suitable fits 
(0.76 $<$ $\chi^{2}_\nu$ $<$ 1.36) to  all  
spectra, with temperatures ranging between 18 and 43 keV. The
power law model failed to fit most of the spectra. The weighted mean of the
OTTB temperatures for this sample of 32 events is $20.8 \pm 0.2$ keV.

To increase our burst sample we performed an off-line search for untriggered 
BATSE events from SGR 1806-20 using a method explained in detail by 
Woods et al. (1999a).
Figure 1 shows the overall BATSE burst activity history of SGR 1806-20.
We limited our search during active phases of the source.
We found, in addition to the 74 triggered events, 
42 untriggered bursts during the time intervals 1993 September 13 -- 
1993 November 20 and 1995 September 7 -- 1999 October 26.
Of these 116 events, 
111 events (triggered and untriggered) had DISCLA data and were sufficiently 
intense to allow spectral fitting. 
Because of the long DISCLA data integration time (1.024 s) compared to
typical SGR burst durations ($\sim$ 0.1 s), we could estimate only the 
fluence for each event. We fit the background-subtracted source spectrum to an 
OTTB model with a fixed kT of 20.8 keV, a reasonable choice considering the 
fairly narrow kT distribution of the triggered bursts derived above.
We find that the burst fluences  
range between $1.4 \times 10^{-8}$ and
$4.3 \times 10^{-6}$ ergs cm$^{-2}$. For a distance to SGR
1806-20 of 14.5 kpc (Corbel et al.~1997)\markcite{cor97}, and assuming
isotropic emission, the corresponding  
energy range is $3.5 \times 10^{38}$ -- $1.1 \times 10^{41}$ ergs.
In comparison, the energies of SGR 1900+14 bursts seen with BATSE
range between $1.1 \times 10^{38}$ -- $1.5 \times 10^{41}$ ergs 
({G\"o\u{g}\"u\c{s}} et al. ~1999) and those of SGR 1627-41 between
$8.0 \times 10^{37}$ -- $5.5 \times 10^{41}$ ergs (Woods et al. 1999b).

\section{RXTE Observations}

We performed 13 pointed observations of SGR 1806-20 with the RXTE/PCA,
for a total effective exposure time of $\sim$ 141 ks between 1996 November 5 
and 18. We searched PCA Standard 1 data (2-60 keV) with 0.125 s time resolution 
for bursts using the following procedure.
For each 0.125 s bin, we estimated a background count rate by 
fitting a first order polynomial
to 5 s of data before and after each bin with a 3 s gap between the bin
searched and the background intervals.
Bins with count rates exceeding 125 counts/0.125 s were assumed to 
include burst emission and were excluded from the background intervals. 
A burst was defined as any continuous set of bins with count rates  
above 5.5 $\sigma$ of the estimated background.
For the typical PCA count rate of 12 $-$ 18 counts/0.125 s in this energy
band, 5.5 $\sigma$ level corresponds to $\sim$ 20 $-$ 25 counts in a 0.125 s 
bin.
We found 290 events and measured the count fluence of each burst by simply 
integrating the background-subtracted counts over the bins covering the event.

To compare the integrated count fluences obtained with the PCA to the
BATSE fluences, we determined a conversion factor between the two as follows. 
First, we searched for bursts observed with both
instruments and found 8 such events (5 of which had triggered BATSE). 
Assuming a constant OTTB model as described in Section 2, we estimated the
fluence of these bursts.
We then computed the
ratio of the BATSE fluence to the PCA counts of each common event.
These ratios fall within a fairly narrow range 
($3.5 \times 10^{-12}$ and $8.1 \times 10^{-12}$ ergs
cm$^{-2}$counts$^{-1}$).
Their weighted mean is $5.5 \times 10^{-12}$ ergs cm$^{-2}$counts$^{-1}$
with a standard deviation,  $\sigma$ = $1.3 \times 10^{-12}$ ergs cm$^{-2}$
counts$^{-1}$.
The mean is very close to the one estimated for SGR 1900+14 
({G\"o\u{g}\"u\c{s}} et al. ~1999) and consistent with the idea that SGR bursts 
have a similar spectral shape. 
Using this conversion factor, we find that the
fluences of the PCA bursts range from $1.2 \times 10^{-10}$ to 
$1.9 \times 10^{-7}$ ergs cm$^{-2}$ and the burst energies range from $3.0 
\times 10^{36}$ to $4.9 \times 10^{39}$ ergs.

\section{ICE Observations}

From 1978 to 1986 the Los Alamos GRB detector on board ICE satellite 
(Anderson et al. 1978) almost continuously observed 
the Galactic center region within which SGR
1806-20 is located. It detected
134 bursts
from the source between 1979 January 7 and 1984 June 8 (Laros et al. ~1987,
~1990; Ulmer et al. ~1993). Combining observational details given 
by Ulmer et al. (1993) and energy spectral information obtained by OTTB fits 
to bursts (at energies E $>$ 30 keV)
given by Fenimore et al. (1994) and Atteia et al. (1987),
we estimate that the ICE burst fluences range form  
$1.5 \times 10^{-8}$ to $6.5 \times 10^{-6}$ ergs cm$^{-2}$ and 
their corresponding isotropic energies are between $3.6 \times 10^{38}$ and 
$1.6 \times 10^{41}$ ergs.

\section{Statistical Data Analysis and Results}

From the previous 3 sections, we clearly see that the BATSE and ICE detection
sensitivities are quite similar, with PCA extending the logN$-$logP distribution
to lower values. We now combine all data bases to a common set,
enabling several statistical analyses. 

{\it{(i) Burst fluence distributions :}}~~
To eliminate systematic effects  
due to low count statistics or binning,
we have employed the maximum likelihood technique to fit the unbinned
burst fluences. A power law fit to 92 BATSE fluence values between 
$5.0 \times10^{-8}$ and $4.3 \times 10^{-6}$ ergs cm$^{-2}$ yields a power
law exponent, $\gamma$ = $1.76 \pm 0.17$ (68$\%$ confidence level). 
Bursts with fluences below $5.0 \times10^{-8}$ ergs cm$^{-2}$ were excluded
to avoid undersampling effects due to lower detection efficiency.
Figure 2 shows the BATSE fluences binned into equally spaced logarithmic 
steps (filled circles).
Similarly, we fit the 266 PCA fluence values between
$1.7 \times 10^{-10}$ and $1.9 \times 10^{-7}$ ergs cm$^{-2}$ 
to a power law model 
and obtain a best fit exponent value of 1.43 $\pm$ 0.06 (see Fig 2,
diamonds for PCA).
Finally, the 113 ICE fluences between $1.8 \times 10^{-7}$ and $6.5 \times 
10^{-6}$ ergs cm$^{-2}$ yield $\gamma$ = 1.67 $\pm$ 0.15 (see Fig 2 squares for
ICE).
We find that the power law indices obtained for BATSE and 
ICE agree well with each other, while the index obtained from 
PCA is marginally lower.

We fit the ICE fluences to a power law $\times$ exponential 
model and to a broken power law model to search for 
evidence of a turnover claimed by Cheng et al. (1996).
Neither model provides a statistically significant improvement 
over a single power law fit.
It is important to note that there is no evidence of a high energy cut-off
or a break in the energy distribution (see Fig 2).

{\it{(ii) Waiting times distribution:}}~~ 
To measure the waiting times between successive SGR 1806-20 bursts,
we identified 22 RXTE observation windows containing two or more bursts without
any gaps. We
then determined 262 recurrence interval times $\Delta$T (i.e. time difference
between successive bursts).
Figure 3 shows a histogram of the $\Delta$Ts, which range from
0.25 to 1655 s.
We have fit the ($\Delta$T)-distribution to a lognormal function and found
a peak at $\sim$ 97 s (with $\sigma \sim$ 3.6). 
This fit does not include waiting times less than 3 s to avoid contribution of
double peaked events in which the second peak appears shortly 
($\sim$ 0.25$-$3 s) after the first one.
To correct for biases due to the RXTE observation window ($\sim$ 3000 s),
we performed extensive numerical simulations and found that the intrinsic
peak of the distribution should be at $\sim$ 103 s.
Note that the observation windows with no bursts may represent 
a long-waiting-time tail which is additional to the lognormal distribution.

To investigate the relation between the waiting time till the next
burst ($\Delta$$T^{+}$) and the intensity of each burst, we divided the 290 
events sample into 6 intensity intervals, each of which contains
approximately 50 events. We fit the $\Delta$$T^{+}$-distribution 
also to a lognormal distribution and determined each peak
mean-{$\Delta$$T^{+}$} (which range from 82 s to 148 s) and 
the mean counts for each of the 6 groups. 
We show in Figure 4 (a) that there is no
correlation between $\Delta$$T^{+}$ and the total burst counts (the Spearman
rank-order correlation coefficient, $\rho$ = $-$0.2 with a probability that
this correlation occurs in a random data set, P = 0.70). 
Similarly, we investigated the relation between the elapsed times since the
previous burst ($\Delta$$T^{-}$) and the intensity of the bursts.
We find that mean-$\Delta$$T^{-}$ extends from 77 s to 120 s. 
Figure 4 (b) shows that there is also no correlation between 
mean-{$\Delta$$T^{-}$} and the burst counts ($\rho$ = $0.4$, P = 0.46).

\section{Discussion}

The fluence distributions of the SGR 1806-20 bursts seen with ICE and 
BATSE
are well described by single power laws with indices 1.67 $\pm$ 0.15 and 
1.76 $\pm$
0.17, respectively, while RXTE bursts have an index of 1.43 $\pm$ 0.06.
These indices are similar to those found for SGR 1900+14 
(1.66, {G\"o\u{g}\"u\c{s}} et al. 1999) and SGR 1627-41 (1.62, 
Woods et al. 1999b).
The ICE and BATSE values are consistent with one another, over nearly the same
energy range but at different epochs. This suggests that SGR event fluence 
distributions
may not vary greatly in time, therefore, we combine the ICE and BATSE values to 
calculate a ``high-energy'' 
index, $\gamma$ = 1.71 $\pm$ 0.11.   
The difference between the ``low-energy'' (RXTE) index and the ``high-energy''
index is insignificant ($\sim$ 2.3 $\sigma$); more ``high-energy'' data are
needed to determine whether there is a break in the distribution.

Power law energy distributions have also been found for earthquakes with  
$\gamma$ = 1.4 to 1.8 (Gutenberg \& Richter 1956; Chen et al. 1991; 
Lay \& Wallace 1995), 
and solar flares, $\gamma$ = 1.53 to 1.73 (Crosby et al. 1993, Lu et al.
1993).
This is a typical behavior seen in self-organized critical systems. 
The concept of
self-organized criticality (Bak, Tang \& Wiesenfeld 1988) states that 
sub-systems
self-organize due to some driving force to a critical state at which a slight
perturbation can cause a chain reaction of any size within the system. 
SGR power law fluence distributions, along with a lognormal waiting time 
distribution 
support the idea that systems responsible for SGR bursts are in a
state of self-organized criticality. 
We believe that in SGRs, the critical systems are neutron star crusts
strained by evolving magnetic stresses (cf. Thompson \& Duncan ~1995).

Cheng et al. (1996) suggested that there is a high energy cut-off in the
cumulative energy distribution of SGR 1806-20 bursts seen by ICE. In a
cumulative energy distribution, the values of neighboring points are correlated,
consequently, judging the significance of apparent deviations is very difficult.
For these reasons we used a maximum likelihood fitting technique and displayed
the differential energy distributions (e.g Fig.2). We find no evidence for a
high-energy cut-off in the ICE data of SGR 1806-20 up to burst energies
$\sim 10^{41}$ ergs.
It should be noted, however,
that a high energy cut-off or turnover must exist because otherwise the total 
energy diverges.

The distribution of waiting times of SGR 1806-20 bursts observed with RXTE
is well described by a lognormal function, similar to that found by Hurley et
al. (1994) for the bursts seen with ICE. The waiting times of the RXTE events 
are on
average shorter than the ones observed with ICE, maybe due to different burst
active phase of the source or to instrumental sensitivity (the PCA is 
more
sensitive to weaker bursts than ICE, and the system displayed plenty of weaker
bursts as well as strong ones in 1996), or combination of both. 
Recently  G\"o\u{g}\"u\c{s} et al. (1999) showed that the recurrence time 
distribution of SGR
1900+14 bursts observed with RXTE is also a lognormal function which peaks at
$\sim$ 49 s. 
The lack of any correlation between
the intensity and the waiting time until the next burst
agrees well with the results of ICE observations of SGR 1806-20 (Laros et
al. 1987). This behavior, also seen in SGR 1900+14 (G\"o\u{g}\"u\c{s} et al. 
1999) confirms
that the physical mechanism responsible for SGR bursts is different from systems
where accretion-powered outbursts take place (e.g. the Rapid Burster, 
Lewin et al. 1976, and the Bursting Pulsar, Kouveliotou et al. 1996)

The burst activity of SGR 1806-20 over the last three years is considerably 
different from that of SGR 1900+14. After a long period with almost no bursts, 
BATSE recorded 200 bursts from SGR 1900+14 between 1998 May and 1999 January,
with a remarkably low activity thereafter.
On the other hand, after SGR 1806-20 reactivated in 1996, it
continued bursting on a lower rate, with 18 bursts in 1997, 32 in 1998
and 18 in 1999 through October. 
The latest RXTE observations of SGR 1806-20 in 1999 August revealed 
that smaller scale bursts are still occurring occasionally in this system,
whereas contemporaneous RXTE observations of SGR 1900+14 do not show 
burst activity of any size.
This continuation of burst activity may prevent 
the deposition of very large amounts of stress in the crust. 
Therefore, in SGR 1806-20 it may be
less likely to expect, in the near future, a giant flare from this source,
as the ones seen on 1979 March 5 from SGR 0526-66 (Mazetz et al. 1979)
and on 1998 August 27 from SGR 1900+14 (Hurley et al. 1999).

\acknowledgments

We are grateful to the referee, Dr. David Palmer for his very constructive 
comments. 
We acknowledge support from NASA grant NAG5-3674 (E.G., J.v.P.)
the cooperative agreement NCC 8-65 (P.M.W.); 
NASA grants NAG5-7787 and NAG5-7849 (C.K.);
Texas Advanced
Research Project grant ARP-028 and NASA grant NAG5-8381 (R.C.D.).

\newpage

\begin{figure}[h]
\plotone{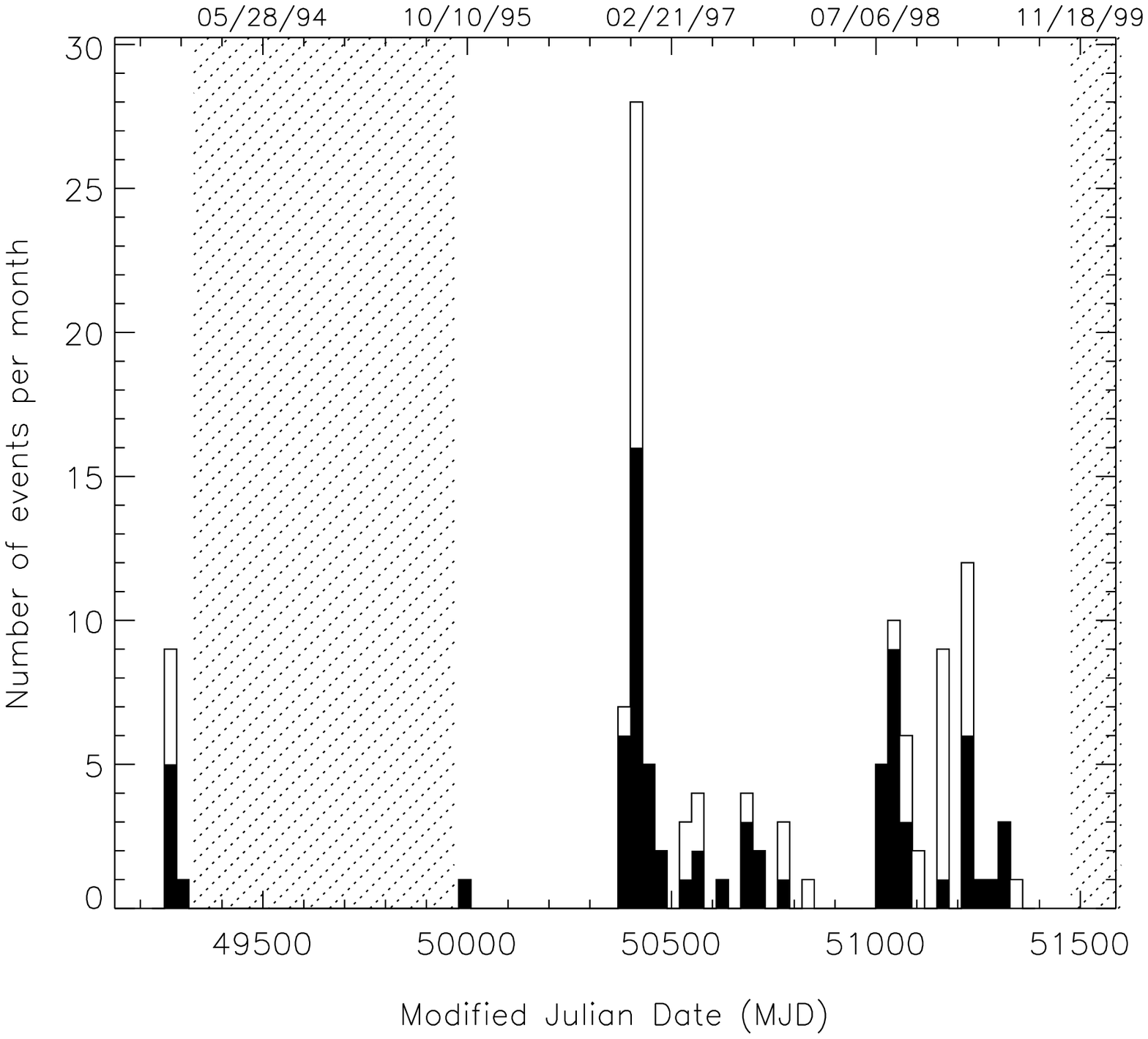}
\caption{Plot of activity history of SGR 1806-20 as seen with BATSE.
Shaded
regions denote the time intervals within which the off-line untriggered
burst
search was not performed. The filled parts illustrate the events within
each time
bin which led to an on-board trigger.}
\label{onebarrel}
\end{figure}

\newpage

\begin{figure}[h]
\plotone{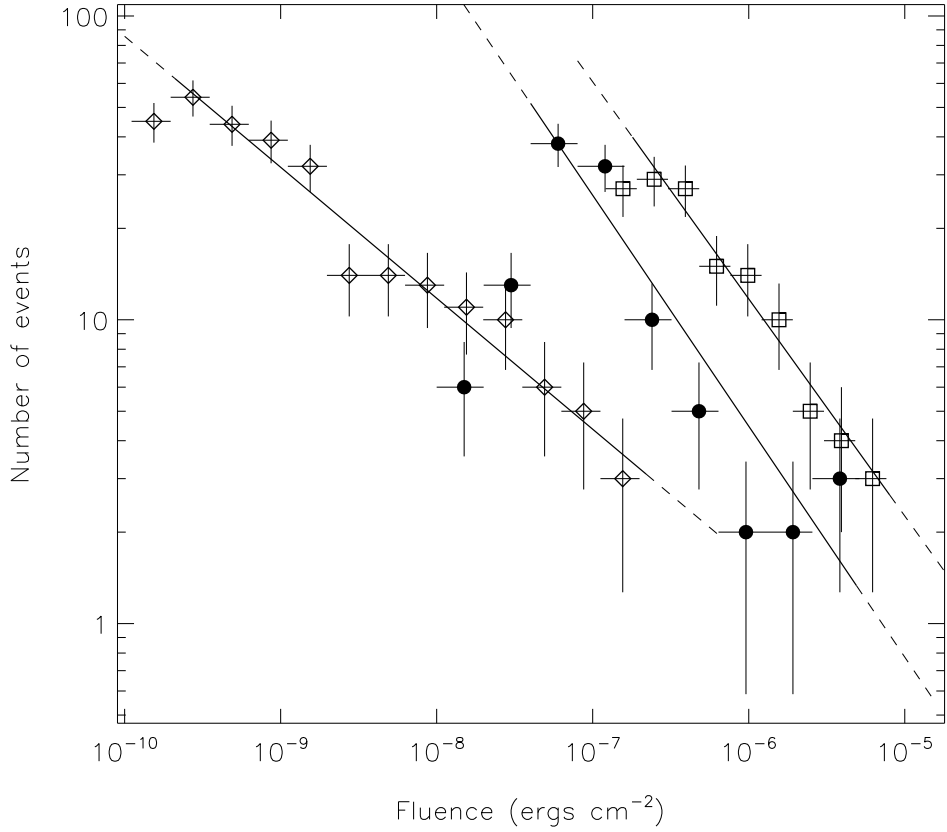}
\caption{Differential fluence distributions of SGR 1806-20 bursts as seen
by
RXTE (diamonds), BATSE (filled circles) and ICE (squares). The lines are
obtained fitting a power law model with the maximum likelihood technique.
The solid lines show the intervals used in the fit and the dashed lines
are the extrapolations of each model.}
\label{onebarrel}
\end{figure}

\newpage

\begin{figure}[h]
\plotone{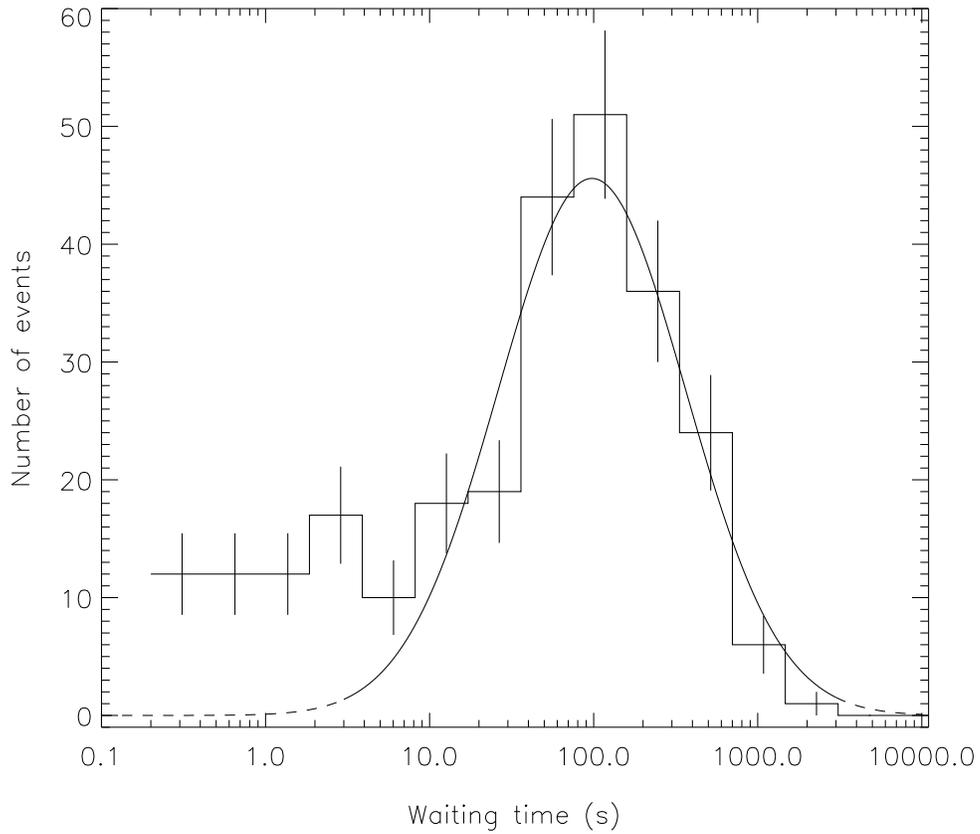}
\caption{Histogram of the waiting times, $\Delta$T, between successive RXTE PCA
bursts from SGR 1806-20. The line shows the best fit lognormal function.
The solid portion of the line indicates the data used in the fit. The excess
of short intervals above the model is due to the double peaked events as
explained in the text.}
\label{onebarrel}
\end{figure}


\begin{figure}[h]
\plotone{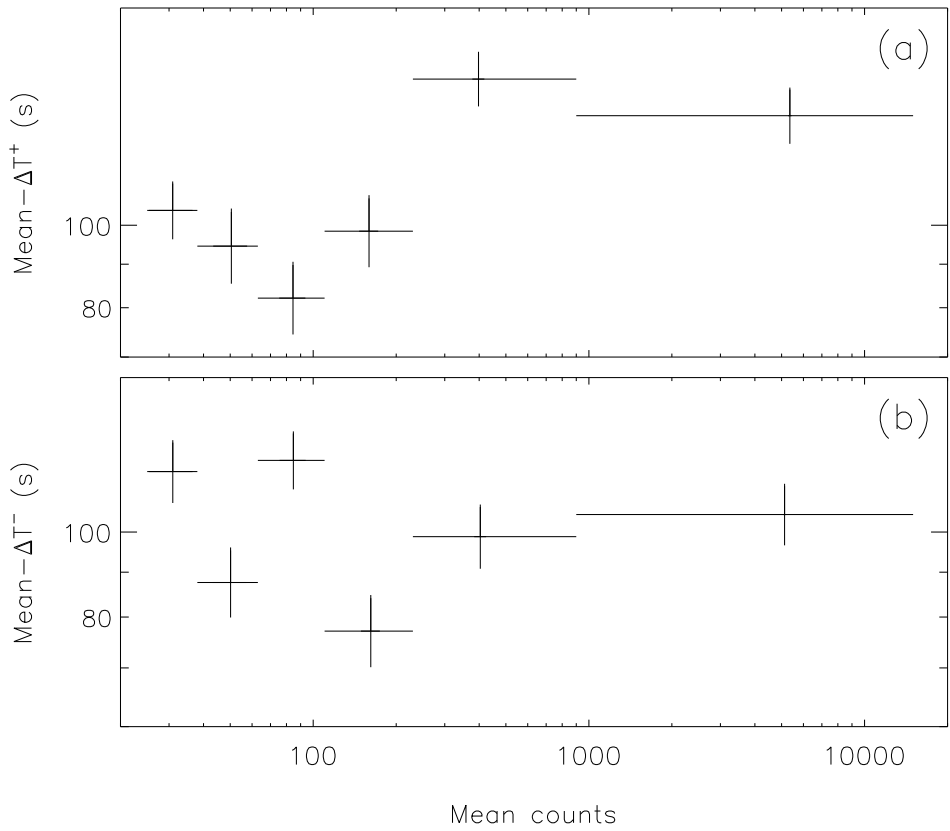}
\caption{(a) Plot of lognormal mean waiting times till the next burst
($\Delta$$T^{+}$)
vs mean total counts. No correlation is seen ($\rho$ = $-$0.2, P=0.70);
(b) The plot of lognormal mean elapsed
times since the previous burst ($\Delta$$T^{-}$) vs mean counts
does not show any correlation either ($\rho$ = 0.4, P=0.46).} 
\label{onebarrel}
\end{figure}

\end{document}